\documentclass[11pt,a4paper]{article}
\usepackage{jheppub_kim}
 \usepackage{slashed}
\usepackage{color}
\usepackage{epsfig}
\usepackage{amsmath}
\usepackage{graphicx}
 \usepackage{graphics}
\usepackage{hyperref}

\def\b{\begin{equation}} \def\e{\end{equation}}

\def\ba{\begin{array}} \def\ea{\end{array}}

\def\bee{\begin{enumerate}}
	\def\eee{\end{enumerate}}

\def\b{\begin{equation}}
\def\e{\end{equation}}
\def\bee{\begin{enumerate}}
	\def\eee{\end{enumerate}}

 \usepackage[hang,nooneline,scriptsize]{subfigure}

\begin{document}
 	\title{Exponential corrected thermodynamics of Born-Infeld BTZ black holes in massive gravity}
\author[a,b,c]{B. Pourhassan,}

\author[d]{M. Dehghani,}

\author[e,f,a,*]{S. Upadhyay,
\note[*]{Visiting Associate   at Inter-University Centre for Astronomy and Astrophysics (IUCAA),
Pune, Maharashtra 411007, India.}}
\author[g]{\.{I}. Sakall{\i},}
\author[h]{and  D. V. Singh}

\affiliation[a] {School of Physics, Damghan University, Damghan, 3671641167, Iran.}
\affiliation[b] {Physics Department, Istanbul Technical University, Istanbul 34469, Turkey.}
\affiliation[c] {Canadian Quantum Research Center 204-3002 32 Avenue Vernon, British Columbia V1T 2L7 Canada.}
\affiliation[d] {Department of Physics, Razi University, Kermanshah, Iran.}
\affiliation[e] {Department of Physics, K.L.S. College,  Nawada, Bihar  805110, India.}
\affiliation[f] {Department of Physics, Magadh University, Bodh Gaya, Bihar 824234, India.}
\affiliation[g] {Department of Physics, Eastern Mediterranean University, Famagusta 99628, North Cyprus via Mersin 10, Turkey.}
\affiliation[h] {Department of Physics, Institute of applied Science and Humanities, GLA University, Mathura 281406 Uttar Pradesh,
India.}
\emailAdd{b.pourhassan@du.ac.ir}
\emailAdd{m.dehghani@razi.ac.ir}
\emailAdd{sudhakerupadhyay@gmail.com; sudhaker@associates.iucaa.in}
\emailAdd{izzet.sakalli@emu.edu.tr}
\emailAdd{veerdsingh@gmail.com}

\abstract{It is known that entropy of black hole gets correction at quantum level. Universally, these corrections are logarithmic and exponential in nature. We analyze the impacts of these quantum corrections on thermodynamics of Born-Infeld BTZ black hole in massive gravity by considering both such kinds of correction. We do comparative analysis of corrected thermodynamics with their equilibrium values. Here, we find that the exponential correction yields to the second point of the first order phase transition. Also, quantum correction effects significantly on the Helmholtz free energy of larger black holes. We study the equation of state for the exponential corrected black hole to obtain a leading order virial expansion.}


\maketitle

\section{ {Introduction}}
Even though  three  dimensional gravity  has  no Newtonian limit, it has always been useful for conceptual issues. This was further supported by  discovery of the three dimensional Banados--Teitelboim--Zanelli  (BTZ)   black hole solution \cite{1,2,3}. This discovery of   three dimensional  BTZ black holes  really helped in recent developments in gravity, gauge and string theory.  The behavior of geodesics \cite{04,05}   and the propagation of strings in a BTZ background \cite{06}  had already been discussed.  In contrast to Schwarzschild and Kerr solutions,  the BTZ black hole  is asymptotically anti-de Sitter and has no curvature singularity at center. Due to  asymptotically AdS in nature of such BTZ black hole solution, it strengthen the idea of  AdS/CFT correspondence. BTZ black holes are widely studies \cite{4,5,6}. The noncommutativity and scattering of massless planar scalar waves had been discussed in the context of   BTZ black hole \cite{7}.

On the other hand,  there exist a lot of reasoning to modify the Einstein gravity.
The idea to include massive gravitons is one of them which explains the current acceleration of the universe without considering a cosmological constant \cite{8,9}.  The solution of
hierarchy problem  hints the existence of massive modes \cite{10}.  Massive gravity gets relevance in astrophysics also as neutron stars can have mass more than $3M_\odot$ in massive gravity scenario \cite{11}. The  massive graviton affects the gravitational waves   also \cite{12}.  Various black hole solution have been studied in four- and higher-dimensional massive gravity \cite{jhep, prd, mpla}. One can not ignore the possibility to explore BTZ black hole
in massive gravity. An AdS charged BTZ black hole   in a massive  gravity has been explored \cite{hendihep}. Here,  thermodynamics and phase structure have also been discussed in both grand canonical and canonical ensembles \cite{hendihep}.

Bekenstein, profoundly, interpreted the area of the event horizon of a black hole $(\mathcal{A})$ as its entropy  $(S_{0})$ \cite{bak}. This is given by
\b
S_{0}=\frac{\mathcal{A}}{4l_{p}^{2}},\label{s0}
\e
where  $l_{p}$  is the Planck length. This relationship more or less
agree in the circumstances when black holes are much larger than the Planck scale. However, for relatively smaller black hole this relation needs correction. For such black holes, thermal fluctuations around equilibrium
becomes significant and correct the black hole entropy. There are several ways to obtain such corrections. Recently, it has been suggested that correction terms are universal and have the following approximate shape \cite{2007.15401},
\begin{equation}
S=S_{0}+\alpha \ln{S_{0}}+\frac{\gamma}{S_{0}}+ \omega e^{-S_{0}}+...\label{ss}
\end{equation}
where dots denote higher-order corrections.
Here, we notice that  at first-order entropy gets logarithmic correction  as confirmed by microstate counting in string theory as well as loop quantum gravity \cite{PLB}.
However,  exponential corrected (EC) term occurs when microstate counting
is done for quantum states  on the horizon only \cite{2007.15401}.
The  phase transition and thermodynamics  of BTZ black holes
through the Landau-Lifshitz theory has been given in Ref.  \cite{70}. The
effects of  entropy  correction in the noncommutative BTZ black
holes is presented in Ref. \cite{71}.
Thermal fluctuations of charged black holes in  massive gravity is presented in Ref. \cite{sudha}. The effects of corrected entropy on thermodynamics of various black holes are studied \cite{sudha1,sudha2,sudha3,sudha4,sudha5,sudha6}. Now, the main goal of this paper is to study the exponential correction of the Born-Infeld BTZ black holes in massive gravity thermodynamics.

In  GR  and its
extensions, nonlinear electrodynamics  plays a
crucial role. For example,   it produces many interesting geometries such as regular black holes. The nonlinear electrodynamics is a direct generalization of the Maxwell electrodynamics originated by  Born and Infeld (BI)  in order to remove the central singularity of a point charge \cite{be}.   BI terms also appears in
superstrings scenario more naturally  \cite{08,09}.   Black holes with BI term    are quite  important in astrophysical observations  \cite{010,011,012,013}.   Bardeen  was first who proposed  regular black hole  \cite{014} and   it was found that  regular black hole with BI term only describes the   black hole formation from an initial vacuum region.  Albeit significant progress made on the subject, quantum effects   on the thermodynamics of BI BTZ black hole in massive gravity are still unstudied. This proves  us an opportunity to shed light on this. The BTZ black hole is a solution of the Einstein field equations in three dimensions that describes a black hole with negative cosmological constant. The Born-Infeld theory is a modified theory of electromagnetism that is characterized by a non-linear dependence of the electromagnetic field on the electric and magnetic fields. The combination of these two concepts leads to the concept of a quantum Born-Infeld BTZ black hole, which is a black hole that is characterized by both quantum and non-linear electromagnetic effects.
In the context of massive gravity, the thermodynamics of quantum Born-Infeld BTZ black holes can be studied by considering the effects of the mass term for the graviton on the properties of the black hole. This can include the effects on the black hole temperature, entropy, and other thermodynamic quantities.
It is important to note that the thermodynamics of quantum Born-Infeld BTZ black holes in massive gravity is still a highly theoretical and complex topic that is not yet fully understood. Further research is needed to fully understand the thermodynamics of these objects and their potential implications for our understanding of gravity and the fundamental nature of spacetime. These are motivations behind present work.

The paper is presented systematically in following manner. In Sec. \ref{sec2}, we discuss
the singular solution of BI massive gravity and their equilibrium thermodynamics.
The quantum corrected thermodynamics of such black hole attributed by logarithmic corrected (LC) entropy is presented in
section \ref{sec3}. The corrections in thermodynamics of this black hole attributed by EC entropy is discussed in section \ref{sec4}. Finally, results of paper is summarized in
the last section.

\section{ { BI BTZ black holes in massive gravity }}\label{sec2}
In this section, we shed light on the  gravitational  field equations and  singular solution of the three-dimensional massive gravity in presence of
nonlinear electrodynamics. We also review its thermal properties.
\subsection{Gravitational field equations and  singular solution}
In this section, we shed light on particular massive BTZ black hole solution \cite{MBTZ} in presence of BI source.
Let us begin by writing the action of three-dimensional massive gravity in presence of nonlinear electrodynamics
\cite{ hendiplb,3dmpl}
\b I=\frac{1}{16\pi}\int\sqrt{-g}\left[{ {R}}-2\Lambda+m_G^2\sum_{i=1}^4c_i{\cal{U}}_i(g,f)+L({\cal{F}})\right]d^3x,\label{1}
 \e
where $\Lambda $ is the cosmological constant, $m_G$ is the constant parameter  related to the graviton mass, coefficients $c_i$ are some constant, and $f$  is  fixed symmetric tensor known as reference metric. Here, symmetric polynomials ${\cal{U}}_i$ are given by \cite{cai, hendiprd}
\begin{eqnarray}
{\cal{U}}_1&=& [{\cal{K}}],\nonumber\\
{\cal{U}}_2&=&[{\cal{K}}]^2-[{\cal{K}}^2],\nonumber\\
{\cal{U}}_3&=&[{\cal{K}}]^3-3[{\cal{K}}][{\cal{K}}^2]+2[{\cal{K}}^3],\nonumber\\
{\cal{U}}_4&=&[{\cal{K}}]^4-6[{\cal{K}}]^2[{\cal{K}}^2]+8[{\cal{K}}^3][{\cal{K}}]+3[{\cal{K}}^2]^2-6[{\cal{K}}^4],\label{2}
\end{eqnarray}
where  $[{\cal{K}}]\equiv{\cal{K}}^a_{\ a}$ is eigenvalues of the $3\times 3$ matrix ${\cal{K}}^a_{\ b}=\sqrt{g^{ac}f_{cb}}$. However,  $L({\cal{F}})$  refers to the Lagrangian of nonlinear electrodynamics expressed in terms of Maxwell's invariant ${\cal{F}}=F^{ab}F_{ab}$,  where field-strength tensor of  vector field $A_a$
has the following form: $F_{ab}=\partial_a A_b-\partial_b A_a$.
Specifically, here, we are interested in following BI type  Lagrangian of nonlinear electrodynamics \cite{2012}:
\b L({\cal{F}})=
	4b^2\left(1-\sqrt{1+\frac{{\cal{F}}}{2b^2}}\right),\label{3}
	\e
where   $b$ is a constant parameter called as   nonlinearity parameter. In the large $b$  limit ($b\rightarrow\infty$), the Lagrangian (\ref{3}) corresponds to the   Maxwell's classical electrodynamics.

For the given action (\ref{1}),  the field equations
corresponding to gravitational field $g_{ab}$ and vector field $A_b$
are given, respectively, by \cite{cai, hendiprd}
\b{ {R}}_{ab}-\frac{1}{2}{ {R}} g_{ab}+\Lambda g_{ab}+m^2_G
\chi_{ab}= \frac{1}{2}L({\cal{F}})
g_{ab}-2\frac{dL ({\cal{F}})}{d{\cal{F}}}F_{ac}F^{\;\;c}_{b}, \label{4}\e
\b\partial_a\left[\sqrt{-g}\frac{dL ({\cal{F}})}{d{\cal{F}}}F^{ab}\right]=0, \label{5}\e
 where tensor $\chi_{ab}$ related to massive graviton has the following expression \cite{cai}:
\begin{eqnarray}
\chi_{ab}=&-&\frac{c_1}{2}\left({\cal{U}}_1g_{ab}-{\cal{K}}_{ab}\right)-\frac{c_2}{2}\left({\cal{U}}_2g_{ab}-2{\cal{U}}_1{\cal{K}}_{ab}+2{\cal{K}}^2_{ab}\right)\nonumber\\
&-&\frac{c_3}{2}\left({\cal{U}}_3g_{ab}-3{\cal{U}}_2{\cal{K}}_{ab}+6{\cal{U}}_1{\cal{K}}^2_{ab}-6{\cal{K}}^3_{ab}\right)\nonumber\\
&-&\frac{c_4}{2}\left({\cal{U}}_4g_{ab}-4{\cal{U}}_3{\cal{K}}_{ab}+12{\cal{U}}_2{\cal{K}}^2_{ab}-24{\cal{U}}_1{\cal{K}}^3_{ab}+24{\cal{K}}^4_{ab}\right).\label{6}
\end{eqnarray}
In order to have the  spherically symmetric solution  for the field equations (\ref{4}) and (\ref{5}),
 we make an  ansatz for the line element \cite{3drain, 3dn}:
\b
ds^2=-f(r)dt^2+\frac{dr^2}{f(r)}+r^2d\theta^2, \label{7}\e
where the specific form of metric function $f(r)$ will be determined in the presence of nonlinear (BI) electrodynamics.

For the following reference metric  $f_{ab}=\mbox{diag}(0,0,c^2)$, the polynomials introduced by equation (\ref{2}) are simplified to
 \cite{cai,hendihep},
\b
{\cal{U}}_1=\frac{c}{r},\;\;\;\;\;
\mbox{and}\;\;\;\;\;{\cal{U}}_2={\cal{U}}_3={\cal{U}}_4=0, \label{8}\e
where $c$ is a positive constant.

For, these values of polynomials, the components of tensor $\chi_{ab}$    (\ref{6}) identified to
\begin{eqnarray}\label{9}
\chi_{tt}=-\frac{cc_1}{2r}g_{tt}, \
\chi_{rr}=\frac{cc_1}{2r}g_{rr},\ \
\chi_{\theta\theta}=0.
\end{eqnarray}

Noting the facts that non-vanishing components of field-strength tensor are $F_{tr}=-F_{rt}=-\partial_rA_t(r) $ and, thus, Maxwell's invariant can be expressed as
$ {\cal{F}}=-2F^2_{tr} \label{11}$.
Corresponding to relations (\ref{5}) and (\ref{7}), the electromagnetic field equation leads to
\b F_{tr}(r)=
	\frac{q}{r\beta},\;\;\;\;\;\;\;\; \mbox{and} \;\;\;\;\;\;\;\; A_t(r)=-q\ln \left[\frac{r}{2\ell}\left(1+\beta\right) \right].\label{12}\e
Here $q$ refers to an integration constant which is relevant for the black hole charge and $ \beta=\sqrt{1+\frac{q^2}{b^2r^2}}$ \cite{dark}.

Corresponding to  the metric (\ref{7}), the components of gravitational field
equations (\ref{4})  take the following differential forms:
\b
G_{tt}=G_{rr}=\frac{f'(r)}{r}+2\Lambda+4b^2\left(\beta-1\right)-\frac{cc_1m_G^2}{r}=0,\label{15}\e
\b G_{\theta\theta}=f''(r)+2\Lambda+4b^2\left(\beta^{-1}-1 \right)=0. \label{16}\e
Here, we observe that the components of the gravitational field
equations hold  the following relation:
\b
\left(r\frac{d}{dr}+1\right)G_{tt}=G_{\theta\theta}.\label{19}\e
Since the components of field equations are interrelated, we need to solve only one (preferably first-order one)  of equations (\ref{15}) and (\ref{16}).
 This leads to the solution for   metric function as
\b
f(r)= 	-m-\Lambda r^2+m_G^2cc_1r-2b^2r^2\left( \beta-1\right)+q^2-2q^2\ln\left[\frac{r}{2\ell}\left( 1+\beta\right)\right],\label{20}\e
where mass parameter $m$ is a constant of integration.

This is worth to calculate   Ricci  scalar and  Kretschmann scalar  as they play significant role in the study of space-time singularities. Thus, the Ricci scalar  and the Kretschmann scalar for this theory  of gravity are calculated, respectively, by
\begin{eqnarray}
  {R} &=&
	6\Lambda+4b^2\left[\left( \beta^{-1}-1\right)+2\left(\beta-1\right)\right]-\frac{2m_G^2cc_1}{r},\label{22}\\
{ {R}}^{\mu\nu\rho\lambda}{ {R}}_{\mu\nu\rho\lambda}&=&
12\Lambda^2+16b^4\left[\left(
\beta^{-1}-1\right)^2+2\left(\beta-1\right)^2\right]-\frac{8b^2
m_G^2cc_1}{r}\left(\beta-1\right) \nonumber\\
 &+& 16\Lambda b^2\left[\left( \beta^{-1}-1\right)
 +2\left(\beta-1\right)\right]-\frac{8\Lambda m_G^2cc_1}{r}+
 \frac{2(m_G^2cc_1)^2}{r^2}.\label{23}
\end{eqnarray}
It is obvious from Eqs. (\ref{22}) and (\ref{22}) that the Ricci scalar and Kretschmann scalar take finite value for finite  $r$.
The solution described by metric function (\ref{20}) is not a regular solution as the singularity  is  not  a  coordinate one but a essential singularity.  This solution behaves like AdS black holes asymptotically \cite{dark}. Above solutions are already presented by Refs. \cite{dark,hendihep}

\subsection{Equilibrium thermodynamics}
  In order to discuss the equilibrium thermodynamics of this black hole, we first
   compute  the black hole mass $M$ as \cite{3drain},
\b  M=\frac{m}{8}, \label{31}\e
where $m$ can be estimated by vanishing metric function $f(r)|_{r=r_+}=0$ as
\begin{equation}
	 m=-\Lambda r^2+m_G^2cc_1r-2b^2r^2\left( \beta-1\right)+q^2-2q^2\ln\left[\frac{r}{2\ell}\left( 1+\beta\right)\right].
\end{equation}
The   Hawking temperature using surface gravity is calculated  by
\b T=
	\frac{1}{4\pi }\left[m_G^2cc_1-2\Lambda
	r_+-\frac{4q^2}{r_+(1+\beta_+)}\right],\;\;\;\;\; \mbox{with}\;\;\;\;\beta_+=\beta|_{r=r_+}.\label{24}\e
 The equilibrium  Hawking-Bekenstein entropy for the BI BTZ black holes in massive gravity  rainbow is calculated by
\b
S_{0}=\frac{\pi r_+}{2}.\label{28}\e
The electric potential $\Phi(r_+)$ and  conserved electric charge $Q$ are calculated, respectively, by
\begin{eqnarray}
 \Phi(r_+)&=&
	-q\ln\left[\frac{r_+}{2\ell}\left( 1+\beta_+\right)\right],\label{29}
	\nonumber\\
	Q&=&\frac{q}{2}. \label{30}
\end{eqnarray}
With the  above calculated thermodynamical quantities at equilibrium, one can easily check the validity of
first-law of thermodynamics.

\section{Quantum corrected thermodynamics due to the LC entropy}\label{sec3}
Now, we focus to the correction in entropy as confirmed by microstate counting in string theory as well as loop quantum gravity. The  particular case of the  LC entropy (\ref{ss}) is given by \cite{PLB}
\begin{equation}
S^{(\text{LC})}=S_{0}+\alpha \ln{S_{0}},
\end{equation}
where the expression of equilibrium area-law entropy $S_{0}$ is given by equation (\ref{28}), and $\alpha$ is the correction coefficient \cite{EPL}.
This consideration modifies the thermodynamics  for BI black holes in massive gravity theory.   From equation (\ref{ss}),   the horizon radius
can be expressed in terms of LC entropy as
 \b r_+=\frac{2\alpha}{\pi}W(\eta),\;\;\;\; \mbox{with} \;\;\;\;\eta= \frac{1}{\alpha}\exp\left({\frac{S^{(\text{LC})}}{\alpha}}\right),\label{rr}\e
  where  $W(x)$ is the well-known Lambert function satisfying the equation $W(x)e^{W(x)}=x$ \cite{exp}.

  Now, plugging the value of $r_+$   (\ref{rr}) into Eq. (\ref{31}), we obtain
 \begin{eqnarray}
  M(Q,S^{(\text{LC})})&=&
\frac{-\alpha W(\eta)}{4\pi^2}\left\{2\Lambda \alpha W(\eta)-m_G^2cc_1\pi +4b^2\alpha W(\eta)\left( \beta_{S^{(\text{LC})}}-1\right) \right.\nonumber\\
&-&\left. \frac{2\pi^2Q^2}{\alpha W(\eta)}\left( 1-2 \ln\left[\frac{\alpha W(\eta)}{\pi \ell}\left( 1+\beta_{S^{(\text{LC})}}\right)\right]\right)\right\},
 \end{eqnarray}
where the definition $\beta_{S^{(\text{LC})}}=\sqrt{1+\frac{\pi^2Q^2}{b^2\alpha^2}W^{-2}(\eta)}$ is used.

The Logarithmic quantum corrected  potential and temperature are calculated by
\begin{eqnarray}\label{340}
\Phi^{(\text{LC})}&=&  \left(\frac{\partial M }{\partial Q}\right)_{S^{(\text{LC})}}=  Q\left( 1-2 \ln\left[\frac{\alpha W(\eta)}{\pi \ell}\left( 1+\beta_{S^{(\text{LC})}}\right)\right]\right),\nonumber\\
T^{(\text{LC})}&=&\left(\frac{\partial M}{\partial S^{(\text{LC})}} \right)_Q= \frac{ r_+ \left[m_G^2cc_1-2\Lambda
	r_+-\frac{4q^2}{r_+(1+\beta_+)}\right]}{4(\pi r_++ {2\alpha}) }= \frac{T}{1+\frac{\alpha}{S_0}}.
\end{eqnarray}
The first-law of thermodynamics under the consideration of LC entropy is given by
\begin{equation}\label{first law BI}
dM(Q, S^{(\text{LC})}, \lambda^{(\text{LC})})=T^{(\text{LC})}dS^{(\text{LC})}+\Phi^{(\text{LC})} dQ=TdS^{(\text{LC})}+\Phi^{(\text{LC})} dQ-\alpha d\lambda^{(\text{LC})},
\end{equation}
where,
\begin{equation}
d\lambda^{(\text{LC})}=\frac{T}{S_{0}}dS_{0}=T(r_+)\frac{dr_{+}}{r_{+}}.\label{ABI1}
\end{equation}
Indeed,   $\lambda^{(\text{LC})}$  can be a new thermodynamics parameter responsible for logarithmic correction term. Here, coefficient $\alpha$ can play the role of  its conjugate variable. Hence, the thermodynamic relations (\ref{340}) must be extended as
\begin{eqnarray}
\Phi^{(\text{LC})} &=& \left(\frac{\partial M }{\partial Q}\right)_{S^{(\text{LC})}, \lambda^{(\text{LC})}},\nonumber\\
T^{(\text{LC})} &=& \left(\frac{\partial M}{\partial S^{(\text{LC})}} \right)_{Q, \lambda^{(\text{LC})}},\nonumber\\
\alpha &=& -\left(\frac{\partial M}{\partial \lambda^{(\text{LC})}} \right)_{Q, S^{(\text{LC})}}.\label{34-Ln}
\end{eqnarray}
Certainly,  such relations will help to fix the value of correction coefficient for various horizon radius.  With the help of equations (\ref{24}) and (\ref{ABI1}), the value of
$\lambda^{(\text{LC})}$ is simplified to
\begin{equation}\label{ABI}
\lambda^{(\text{LC})}=\frac{1}{4\pi}\left\lbrace mG^2cc_1 \ln r_++2\left(2b^2-\Lambda \right)r_+-4b \left(br_+\beta_+-q \ln\left[\frac{2}{r_+}+\frac{2b}{q}\beta_+ \right]  \right)    \right\rbrace .
\end{equation}
On the other hand,  Eq.(\ref{ss}) suggests that $S^{(\text{LC})}$ is a function of $S_0$ and $dS^{(\text{LC})}=\left(1+\frac{\alpha}{S_0} \right)dS_0$. This eventually leads to
\begin{equation}\label{first law 2}
dM(Q, S^{(\text{LC})}, \lambda^{(\text{LC})} )=dM(Q, S_0)=T dS_0 +\Phi^{(\text{LC})} dQ.
\end{equation}
which is another form of  the  first-law of thermodynamics.\\

In order to study the effects of LC on the phase transition and  stability of the BI massive black holes,  one requires  the
signature of   specific heat. The  specific heat
of the BI black hole in massive gravity  can be calculated from the following definition:
\b {\cal{C}}^{(\text{LC})}_Q =T\left(\frac{\partial S^{(LC)}}{\partial T}\right)_Q,\e
This yields
\b {\cal{C}}^{(\text{LC})}_Q = \frac{\pi \beta_+[m_G^2cc_1-4b^2r_+(\beta_+-1)-2\Lambda r_+]}{4[2b^2(\beta_+ -1)-\Lambda\beta_+]} \left(1+\frac{2\alpha}{\pi r_+} \right).\label{spe}\e
The vanishing numerator of above expression determines first-order phase transition point. However,  vanishing numerator  of specific heat shows the (divergent) second-order phase transition point.
The expression suggests that second-order phase transition occurs at the center of black hole due to correction term. However, the first-order phase transition occurs at two points of horizon radius.\\

\begin{figure}
\begin{center}$
\begin{array}{cccc}
\includegraphics[width=75 mm]{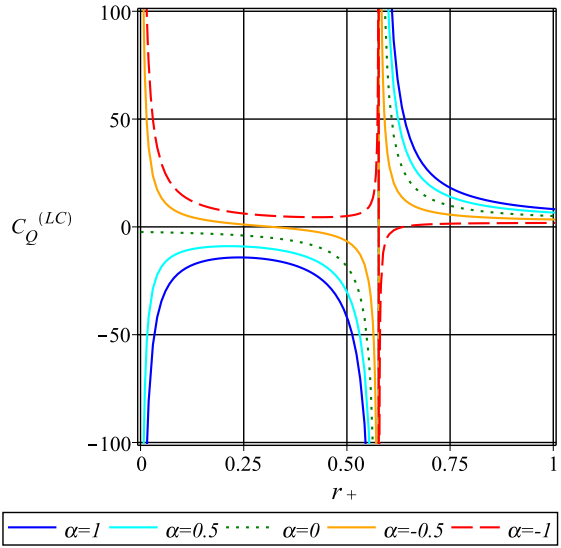}
\end{array}$
\end{center}
\caption{Specific heat ${\cal{C}}^{(\text{LC})}_Q$ in terms of $r_{+}$ for unit values of the model parameters.}
\label{fig1}
\end{figure}

This specific heat (\ref{spe}) can be expressed in terms of equilibrium
specific heat ${\cal{C}}^{(0)}_Q $ as
\b {\cal{C}}^{(\text{LC})}_Q = {\cal{C}}^{(0)}_Q \left(1+\frac{\alpha}{S_0} \right),\e
where
\begin{equation}\label{C0}
{\cal{C}}^{(0)}_Q =\frac{\pi \beta_+[m_G^2cc_1-4b^2r_+(\beta_+-1)-2\Lambda r_+]}{4[2b^2(\beta_+ -1)-\Lambda\beta_+]}
\end{equation}
is uncorrected specific heat corresponding to $\alpha=0$.\\
In Fig. \ref{fig1} we can see typical behavior of ${\cal{C}}^{(\text{LC})}_Q$ in terms of $r_{+}$. Clearly, in the absence of logarithmic correction there is only one point of second-order phase transition located at $r_+=R$. It means that the black holes with horizon radius equal to $R$ undergo second-order phase transition. The black holes with horizon radii greater than $R$ are locally stable while those with horizon radii smaller than $R$ are unstable. The logarithmic correction with positive $\alpha$ (i.e. $\alpha=0.5$ and $\alpha=1$) has no any effect on the stability of BI BTZ black holes. When the logarithmic correction is considered with negative $\alpha$ (i.e. $\alpha=-0.5$ and $\alpha=-1$), in addition to the mentioned second-order phase transition, there is a first-order phase transition too. If  $\alpha=-0.5$ is chosen, the first-order phase transition occurs at $r_+=R_1<R$ and it happens at $r_+=R_2>R$ for $\alpha=-1$. In the case $\alpha=-0.5$, the black holes with horizon radii smaller than $R_1$ and greater than $R$ are stable wile in the case $\alpha=-1$ those with horizon radii smaller than $R$ and greater than $R_2$ remain stable.\\

The Helmholtz free energy is given by,
\begin{equation}\label{FLnBI1}
F^{(\text{LC})}=-\int{S^{(\text{LC})}dT}=-\int{\left( S_{0}+\alpha \ln  S_{0}\right) dT}=F^{(0)}+\alpha \delta F^{(\text{LC})},
\end{equation}

\begin{figure}
\begin{center}$
\begin{array}{cccc}
\includegraphics[width=75 mm]{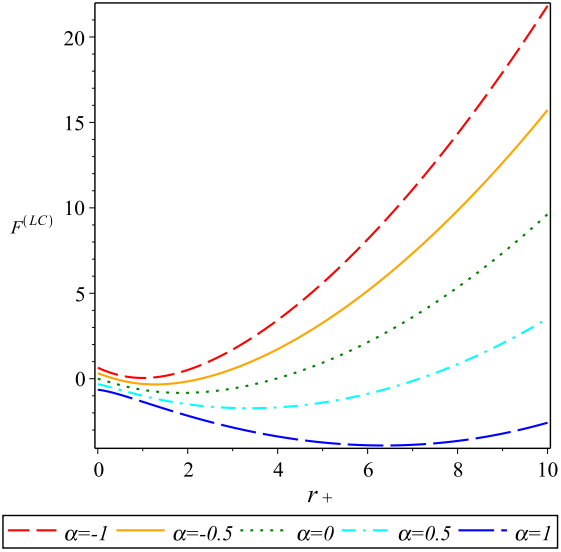}
\end{array}$
\end{center}
\caption{Helmholtz free energy $F^{(\text{LC})}$ in terms of $r_{+}$ for unit values of the model parameters.}
\label{fig2}
\end{figure}

where the equilibrium Helmholtz free energy is given by
\begin{equation}\label{F0BI1}
F^{(0)}=-\int{S_{0}dT}=\frac{1}{8}\left[\Lambda r_+^2-2q^2\left\lbrace 2\ln\left(\frac{\beta_++1}{\beta_+-1} \right)- \frac{1}{\beta_++1}\right\rbrace  \right],
\end{equation}
and the correction term is given by
\begin{eqnarray}
\delta F^{(\text{LC})}&=&-\int {\ln S_{0}\ dT},\nonumber\\
&=&\frac{r_+}{2\pi}\left\lbrace \Lambda\left(\ln r_{+}-1\right)\left(\Lambda-2b^2 \right)-\frac{2qb}{r_+}\ln\left[\frac{2}{r_+}\left(1+\frac{br_+}{q}\beta_+ \right)  \right] \right.\nonumber\\
 &-& \left.  2b^2 \beta_+ \left(\ln r_+ +1\right) \right\rbrace+T\ln\left( \frac{2}{\pi}\right).
\label{F1BI}\end{eqnarray}

In Fig. \ref{fig2}, we can see typical behavior of $F^{(\text{LC})}$ in terms of $r_{+}$. We can see that the logarithmic correction, depending on the correction coefficient, may increase or decrease value of the Helmholtz free energy. In the case of $\alpha=-1$, Helmholtz free energy is completely positive including a positive minimum (see dashed red line of Fig. \ref{fig2}). Solid orange line of Fig. \ref{fig2} represent the case of $\alpha=-\frac{1}{2}$ \cite{EPJC}.\\
The part of the logarithmic corrected thermodynamics presented here is already  presented by the Ref. \cite{MPLA-DP}. However here we succeed to determine the correction coefficient.
\section{Quantum corrected thermodynamics due to the EC entropy}\label{sec4}
In the main part of this paper we would like to consider the exponential correction on the black hole entropy.
As we know, the entropy of black hole gets exponential correction when microstate counting
is performed for quantum states residing on the horizon only \cite{2007.15401}. Here, we consider the entropy perturbation
due to  exponential term only. This is given by \cite{CQG}
\begin{equation}\label{s}
S^{(\text{EC})}=S_{0}+\omega e^{-S_{0}},
\end{equation}
where $\omega$ is the correction coefficient \cite{JHEP1, JHEP2}. This leads to
\b r_+=\frac{2}{\pi}\left[S^{(\text{EC})}-W(\xi)\right],\;\;\;\; \mbox{with} \;\;\;\;\xi= \omega e^{-S^{(\text{EC})}},
\label{exs}\e where, $W(\xi)$ is the well-known Lambert $W$ function.

\begin{figure}
\begin{center}$
\begin{array}{cccc}
\includegraphics[width=75 mm]{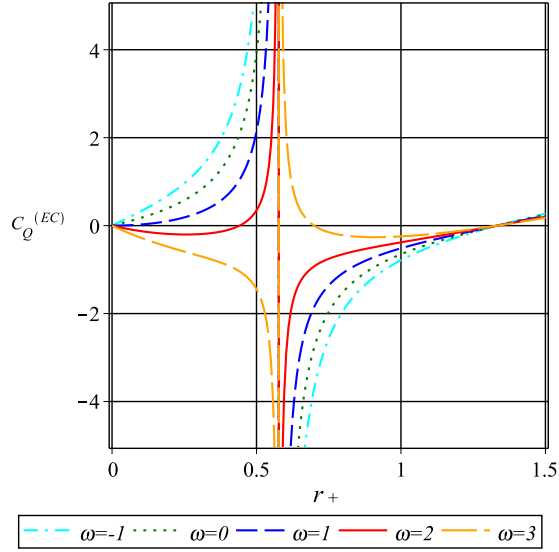}
\end{array}$
\end{center}
\caption{Specific heat ${\cal{C}}^{(\text{EC})}_Q$ in terms of $r_{+}$ for $m_{G}=2$ and unit values of the model parameters.}
\label{fig3}
\end{figure}

 Now, inserting (\ref{exs}) into Eq. (\ref{31}), we obtain
 mass in terms of EC entropy as
\begin{eqnarray}
 M(Q,S^{(\text{EC})})&=&
\frac{W(\xi)-S^{(\text{EC})}}{2\pi^2}\left\{\Lambda\left[S^{(\text{EC})}-W(\xi)\right]+2b^2\left[S^{(\text{EC})}-W(\xi)\right]\left( \omega_{S^{(\text{EC})}}-1\right)\right. \nonumber\\
&-& \left. \frac{\pi}{2} m_G^2cc_1 - \frac{\pi^2Q^2}{ S^{(\text{EC})}-W(\xi) }\left( 1-2 \ln\left[\frac{ [S^{(\text{EC})}-W(\xi) ]\left( 1+\omega_{S^{(\text{EC})}}\right)}{ \ell}\right]\right)\right\},
\end{eqnarray}
where  $\omega_{S^{(\text{EC})}}=\sqrt{1+\frac{\pi^2Q^2}{b^2} [S^{(\text{EC})}-W(\xi) ]^{-2}}$. From the straightforward calculations, we obtain the following expressions:
\begin{eqnarray}\label{34}
\Phi^{(\text{EC})}&=&\left(\frac{\partial M }{\partial Q}\right)_{S^{(\text{EC})}}=Q\left( 1-2 \ln\left[\frac{ [S^{(\text{EC})}-W(\xi) ]\left( 1+\omega_{S^{(\text{EC})}}\right)}{ \ell}\right]\right),\nonumber\\
T^{(\text{EC})}&=&\left(\frac{\partial M}{\partial S^{(\text{EC})}} \right)_Q= \frac{T}{1-\omega e^{-S_0}}.
\end{eqnarray}
By considering EC entropy,  the first-law of thermodynamics  for this black hole is given by
\begin{equation}
dM(Q, S^{(\text{EC})}, \lambda^{(\text{EC})})=T^{(\text{EC})}dS^{(\text{EC})}+\Phi^{(\text{EC})} dQ=TdS^{(\text{EC})}+\Phi^{(\text{EC})} dQ+\omega
d\lambda^{(\text{EC})},\label{BI}
\end{equation}
where $d\lambda^{(\text{EC})}$ has following expression:
\begin{equation}
d\lambda^{(\text{EC})}=Te^{-S_{0}}dS_{0}.\label{1ABI}
\end{equation}
Indeed,   $\lambda^{(\text{EC})}$ is introduced as a new
thermodynamical variable responsible for exponential correction
term with conjugate variable $\omega$. Hence,
the thermodynamic relations (\ref{34}) can be extended as
\begin{eqnarray}
\Phi^{(\text{EC})} &=& \left(\frac{\partial M }{\partial Q}
\right)_{S^{(\text{EC})}, \lambda^{(\text{EC})}},\nonumber\\
T^{(\text{EC})} &=& \left(\frac{\partial M}{\partial S^{(\text{EC})}}
\right)_{Q, \lambda^{(\text{EC})}},\nonumber\\
\omega &=& -\left(\frac{\partial M}{\partial
\lambda^{(\text{EC})}} \right)_{Q, S^{(\text{EC})}}.\label{34-Ln1}
\end{eqnarray}
This relation will be helpful in order to fix the value of EC parameter for
various horizon radius. Finally, Eq. (\ref{1ABI}) simplified to
\begin{equation}\label{ABI2}
\lambda^{(\text{EC})}=-\frac{1}{4\pi}\left[m_G^2cc_1\;e^{-S_0}+
\frac{4}{\pi}\left( 2b^2-\Lambda\right) \left(1+S_0 \right)e^{-
S_0}+\frac{8 b^2}{\pi} \int{\omega_{S_0}e^{-S_0}S_0 dS_0}
 \right].\end{equation}\\
From Eq.(\ref{s}), we note that  $S^{(\text{EC})}$ is a function of $S_0$ and $dS^{(\text{EC})}=\left(1-\omega e^{-S_0}\right)dS_0$. This, eventually, modifies the first-law of thermodynamics (\ref{BI}) as
\begin{equation}\label{first law 3}
dM(Q, S^{(\text{EC})}, \lambda^{(\text{EC})} )=dM(Q, S_0)=T dS_0 +\Phi^{(\text{EC})} dQ.
\end{equation}
Starting from the following definition of the black hole specific heat
\begin{eqnarray}
{\cal{C}}^{(\text{EC})}_Q &=&T\left(\frac{\partial S^{(\text{EC})}}{\partial T}\right)_Q,\nonumber\\
&=&{\cal{C}}^{(0)}_Q \left(1-\omega e^{-S_0} \right),
\end{eqnarray}
where ${\cal{C}}^{(0)}_Q$, as the un-corrected specific heat corresponding to $\omega=0$, is given by the equation (\ref{C0}). In Fig. \ref{fig3} we have shown typical behavior of the specific heat for different values of $\omega$ by letting $m_G=2$. In the absence of correction we see that a second-order phase transition may happen at $r_+=r$ and a first-order one at $r_+=r_1>r$. The BI black holes with the horizon radii in the ranges $r_+<r$ and $r_+>r_1$ are stable. Note that the difference between $\alpha=0$ and $\omega=0$ cases arson from different choices of $m_G$, which reflects the impacts of massive gravitons. In the presence of exponential correction with negative $\omega$ the situation remains unchanged. When $\omega$ is taken positive (i.e. $\omega=2$ and $\omega=3$) stability of the black holes is affected by the exponential quantum correction such that an additional first-order phase transition can occur. In the case $\omega=2$ it appears at $r_+=r_0<r$ and for $\omega=3$ at $r_+=r'$, with $r<r'<r_1$. The black holes with $\omega=2$ are stable for horizon radii in the ranges $r_0<r_+<r$ and $r_+>r_1$, also those with $\omega=2$ and horizon radii in the intervals $r<r_+<r'$ and $r_+>r_1$ prefer thermal stability.

\begin{figure}
\begin{center}$
\begin{array}{cccc}
\includegraphics[width=75 mm]{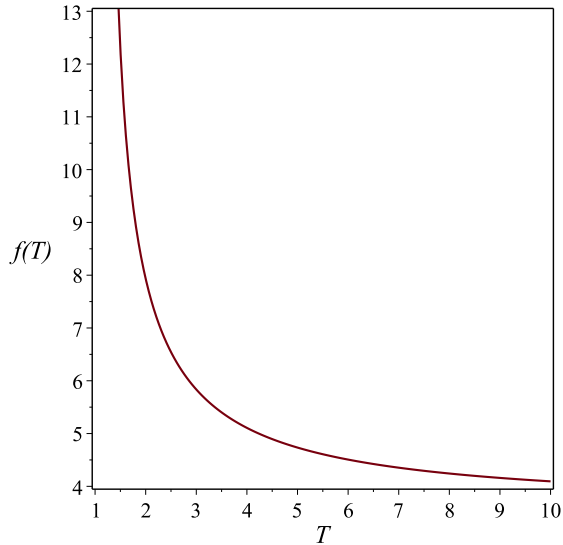}
\end{array}$
\end{center}
\caption{Typical behavior of $f(T)$ in terms of $T$ for unit values of the model parameters.}
\label{fig4}
\end{figure}

The Helmholtz free energy corresponding to EC entropy is derived as
\begin{equation}\label{FLnBI}
F^{(\text{EC})}=-\int{S^{(\text{EC})}dT}=-\int{\left( S_{0}+\omega e^{- S_{0}}\right) dT}=F^{(0)}+\omega \delta F^{(\text{EC})},
\end{equation}
in which $F^{(0)}$ has same form as Eq. (\ref{F0BI1})
and
\begin{equation}\label{F1BI1}
\delta F^{(\text{EC})}=-\int{ e^{- S_{0}}dT}= \frac{e^{- S_{0}}}{\pi^2}\left(\Lambda+2b^2 \right)+ \frac{b^2}{\pi}\int{\frac {e^{- S_{0}} }{\beta_+}}dr_+.
\end{equation}
With this expression, one can check the effects of exponential correction on free energy. Assuming small $b$, one can obtain,
\begin{equation}\label{FLnBI-app}
F^{(\text{EC})}=\frac{r_{+}}{8}+Q\ln{(r_{+})}\frac{3Q}{b^{2}r_{+}^{2}}+\frac{\omega e^{-\frac{\pi r_{+}}{2}}}{\pi^{2}}\left(2b^{2}+\Lambda- \frac{b^{3}(\pi r_{+}+2)}{Q\pi}\right).
\end{equation}
Then, using the thermodynamics relation,
\begin{equation}\label{P}
P=-\left(\frac{\partial F^{(\text{EC})}}{\partial V}\right)_{T},
\end{equation}
one can obtain the pressure as the following expression,
\begin{equation}\label{P1}
P=\frac{\omega b^{2}r_{+}^{3}\left((b^{2}+\frac{\Lambda}{2})Q-\frac{b^{3}r_{+}}{2}\right)e^{-\frac{\pi r_{+}}{2}}-\pi Q\left((b^{2}r_{+}^{2}-6)Q^{2}+\frac{b^{2}r_{+}^{3}}{8}\right)}{2\pi v Q b^{2} r_{+}^{4}},
\end{equation}
where we assumed $V\propto r_{+}^{2}$ (considered $v$ as a proportionality constant). Assuming $Q=\frac{4b}{\pi}$ one can rewrite the equation of state as the following form,
\begin{equation}\label{EoS}
\frac{PV}{T}=1-\frac{B}{V}+f(T),
\end{equation}
where $B=-\frac{6v}{b^{2}}$, and $f(T)$ is a temperature dependent function which we draw its typical behavior in Fig. \ref{fig4}.

We can see that $f(T)$ vanishes at higher temperature and higher temperature means infinitesimal radii where the exponential correction is dominant. Therefore equation of state (\ref{EoS}) at small black hole phase reflects the leading order virial expansion.\\

\section{Conclusion}\label{sec5}

First of all, we have considered the three-dimensional massive gravity coupled with nonlinear electrodynamics and calculated the relate equations of motion. By solving them, we obtained a class of BI BTZ black hole solution. In order to study the space-time singularities, we have estimated the Ricci and Kretschmann scalars which are singular at $r=0$ and, take finite value for finite $r$. This confirms that obtained black hole solution is not a regular solution as the singularity is not due to coordinate one but it is an essential singularity. Asymptotically, this solution behaves like AdS black holes.\\
Furthermore, we have discussed the thermodynamics of such system in equilibrium and, especially entropy, mass, Hawking temperature and electric potential have been calculated. It is well-known that there exists some quantum corrections on the entropy of black holes \cite{2007.15401}.
For instance, the logarithmic correction in entropy has been confirmed by micro-state counting in string theory as well as loop quantum gravity \cite{PLB}. However, the entropy of black hole gets exponential correction when micro-state counting is performed for quantum states residing on the horizon only \cite{2007.15401}. Here, we have first computed  the quantum corrected thermodynamics for the BI BTZ black hole in massive gravity due to the LC entropy. The effects of  quantum correction due to LC entropy on the stability is also studied.  The calculated specific heat   corresponding to LC entropy suggests that in addition to a second-order phase transition there is  first-order phase transition which is due to consideration of LC entropy. However, the logarithmic correction has no effect on the second-order phase transition points. By this we mean that the number and location of second-order phase transition points remain unchanged. We have also found that logarithmic correction affects the Helmholtz free energy significantly for larger black holes. The effects of exponential correction is also studied on the thermodynamics and stability of BI massive BTZ black holes. We have shown that the BI black holes in the presence of massive gravitons exhibit one second-order and one first-order phase transition. So the exponential correction may yield to a second first-order phase transition. Moreover, stability properties of black holes is affected by the mass of graviton. This fact can be seen through comparison of un-corrected specific heats as shown in Figs. 1 and 3 for different values of $m_G$.  Also, the exponential corrected thermodynamics of BI black hole in massive gravity yields to the leading order virial expansion which is not already reported by the ordinary BI black hole \cite{CQG}.


\begin{thebibliography}{a}

\bibitem{1} M.Banados, C. Teitelboim 	nd  J. Zanelli,  Phys. Rev. Lett. 69 (1992) 1849.
\bibitem{2}G. Clement,  Phys. Lett. B 367 (1996) 70.
\bibitem{3}M. Cadoni,  M. Melis and  M.R. Setare,  Class. Quantum Gravit. 25 (2008) 195022.
\bibitem{4}N.-ul-islam, P. A. Ganai and S. Upadhyay, Prog. Theor. Exp. Phys. 103B06 (2019).
\bibitem{04}  N. Cruz, C. Martinez, and L. Pena, Class. Quant. Grav. 11 (1994) 2731.
\bibitem{05}C. Farina, J. Gamboa, and A. J. Segui-Santonja, Class. Quantum Grav. 10 (1993) L193.
\bibitem{06}A. L. Larsen and N. Sanchez, Phys. Rev. D50 (1994) 7493.
\bibitem{5}E. Witten,   Adv. Theor. Math. Phys. 2 (1998) 505.
 \bibitem{6}H.W. Lee, Y.S. Myung and J.Y. Kim,   Phys. Lett. B 466 (1999) 211.
\bibitem{7}M.A. Anacleto, F.A. Brito and E. Passos, Phys. Lett. B 743 (2015) 184.
\bibitem{8}K. Hinterbichler, Rev. Mod. Phys. 84 (2012) 671.
\bibitem{9} C. Deffayet, Phys. Lett. B 502 (2001) 199.
\bibitem{10}G. Dvali, G. Gabadadze, and M. Porrati, Phys. Lett. B 485 (2000) 208.
\bibitem{11}S. H. Hendi, G. H. Bordbar, B. E. Panah, and S. Panahiyan, 	JCAP 07 (2017) 004.
\bibitem{12}A. E. Gumrukcuoglu, S. Kuroyanagi, C. Lin, S. Mukohyama, and N. Tanahashi, Classical Quantum Gravity 29 (2012) 235026.
\bibitem{jhep} S. H. Hendi, B. Eslam Panaha and S. Panahiyan, JHEP 11 (2015) 157.
\bibitem{prd} M. Dehghani, Phys. Rev. D, 106 (2022) 084019.
\bibitem{mpla} M. Dehghani, Mod. Phys. Lett. A, 37 (2022) 2250051.
\bibitem{hendihep}S. H. Hendi, B. Eslam Panaha and S. Panahiyan, JHEP 05 (2016) 029.
\bibitem{bak} J. D. Bekenstein, Phys. Rev. D  7 (1973) 2333.
\bibitem{2007.15401} A. Chatterjee and A. Ghosh, Phys. Rev. Lett.  125 (2020) 041302.
\bibitem{PLB} B. Pourhassan, M. Faizal, Z. Zaz, A. Bhat, Phys. Lett. B 773 (2017) 325.
\bibitem{70}  Z. Y. Tang, C. Y. Zhang, M. K. Zangeneh, B. Wang, and J. Saavedra, arXiv:1610.01744.
 \bibitem{71} M. A. Anacleto, F. A. Brito, E. Passos, A. G. Cavalcanti, and J. Spinelly, arXiv:1510.08444.
\bibitem{sudha}S. Upadhyay, S. H. Hendi, S. Panahiyan and B. E. Panah, Prog. Theor. Exp. Phys. 093E01 (2018).
\bibitem{sudha1}Y. H. Khan, S. Upadhyay, P. A. Ganai, Mod. Phys. Lett. A 36 (2021) 2130023.
\bibitem{sudha2}B. Pourhassan, S. Upadhyay, Eur. Phys. J. Plus 136 (2021) 311.
\bibitem{sudha3}S. Upadhyay, Gen. Rel. Grav. 50 (2018) 128.
\bibitem{sudha4}S. Upadhyay, Phys. Lett. B 775 (2017)  130.
\bibitem{sudha5}S. Upadhyay, B. Pourhassan, Prog. Theor. Exp. Phys. 013B03 (2019).
\bibitem{sudha6}Y. H. Khan, P. A. Ganai, S. Upadhyay, Eur. Phys. J. Plus 135 (2020) 338.
\bibitem{be} M. Born and L. Infeld,  Proc. R. Soc. Lond. 144 (1934)  425.
\bibitem{08}E. S. Fradkin and A. A. Tseytlin, Phys. Lett. B 163 (1985)  123.
\bibitem{09} E. Bergshoeff, E. Sezgin, C. N. Pope  and P. K. Townsend, Phys. Lett. B 188 (1987) 70.
\bibitem{010}W. Javed, R. Babar and A.  Ov\"gun, Phys. Rev. D  100 (2019) 104032.
\bibitem{011} K. Jusufi, A.  Ov\"gun, A. Banerjee and I. Sakallı, Eur. Phys. J. Plus 134 (2019)  428.
\bibitem{012} A.  Ov\"gun, Phys. Lett. B   820 (2021)  136517.
 \bibitem{013} W. Javed, J. Abbas and A. Ov\"gun, Eur. Phys. J. C ¨ 79 (2019)  694.
 \bibitem{014} J.M. Bardeen, in: Conference Proceedings of GR5, Tbilisi, USSR, 1968, p. 174.
\bibitem{MBTZ} S. Chougule, S. Dey,  B. Pourhassan, M. Faizal, Eur. Phys. J. C 78 (2018) 685.
\bibitem{hendiplb}S.H. Hendi, B. Eslam Panah, S. Panahiyan and M. Momennia, Phys. Lett. B  775 (2017) 251.
\bibitem{3dmpl}M. Dehghani, Phys. Lett. B  803 (2020) 135335.
\bibitem{cai} R. G. Cai, Y-P. Hu, Q-Y. Pan and Y-L. Zhang, Phys. Rev. D  91 (2015) 024032.
\bibitem{hendiprd} S.H. Hendi, S. Panahiyan, S. Upadhyay and B. Eslam Panah, Phys. Rev. D  95 (2017) 084036.
\bibitem{2012} S. H. Hendi, JHEP  03 (2012) 065.
\bibitem{3drain} M. Dehghani, Phys. Lett. B  777 (2018) 351.
\bibitem{3dn} M. Dehghani, Phys. Lett. B 799 (2019) 135037.
\bibitem{dark} M. Dehghani, Phys. Dark. Univ.  31 (2021) 100749.
\bibitem{EPL} B. Pourhassan, M. Faizal, EPL  111 (2015) 40006.
\bibitem{exp} M. Dehghani, Phys. Rev. D, 98 (2018) 044008.
\bibitem{Sud} S. Upadhyay, N. Islam,  P. Ganai, J. Holography Applic. Phys. 2 (2022) 25.
\bibitem{EPJC} B. Pourhassan, M. Faizal, U, Debnath, Eur. Phys. J. C 76 (2016) 145.
\bibitem{MPLA-DP} M. Dehghani and B. Pourhassan, Mod. Phys. Lett. A 36 (2021)  2150158.
\bibitem{CQG} B. Pourhassan, M. Dehghani, M. Faizal, S. Dey, Class. Quantum Grav. 38 (2021) 105001.
\bibitem{JHEP1} B. Pourhassan, M. Faizal, JHEP 2021 (2021) 1-18.
\bibitem{JHEP2} B. Pourhassan, S. S. Wani, S. Soroushfar, M. Faizal, JHEP 2021 (2021) 1.

\end{thebibliography}
\end{document}